%% file: conf_06_020_copy.tex
\documentclass[11pt]{article}
\usepackage{graphicx}

\newcommand{\BABARPubYear}    {06}

\newcommand{\BABARConfNumber} {020}
\newcommand{\SLACPubNumber} {12003}

\input pubboard/babarsym
 
\newcommand{\z}{\ensuremath{{\mathsf z}}\xspace}
\newcommand{\Imz}{\ensuremath{\rm Im\, \z}}
\newcommand{\ImzZero}{\ensuremath{\rm Im\, \z_0}}
\newcommand{\ImzOne}{\ensuremath{\rm Im\, \z_1}}
\newcommand{\Rez}{\ensuremath{\rm Re\, \z}}
\newcommand{\dG}{\ensuremath{ \Delta \Gamma }}
\newcommand{\absqop}{\ensuremath{|q/p|}}

\newcommand{\dGRez}{\ensuremath{\dG \times \Rez}}
\newcommand{\dGRezZero}{\ensuremath{\dG \times \Rez_{0}}}
\newcommand{\dGRezOne}{\ensuremath{\dG \times \Rez_{1}}}

\def\dt {\ensuremath{\Delta t}}
\def\dm {\ensuremath{\Delta m}}

\long\def\inst#1{\par\nobreak\kern 4pt\nobreak
    {\it #1}\par\vskip 10pt plus 3pt minus 3pt}

\begin{document}
{\pagestyle{empty}
\begin{flushleft}
\end{flushleft}
\begin{flushright}
\babar-CONF-\BABARPubYear/\BABARConfNumber \\
SLAC-PUB-\SLACPubNumber \\
July 2006 \\
\end{flushright}

\par\vskip 4.7cm

\begin{center}
\Large \bf \boldmath Search for \CPT and Lorentz Violation in \Bz-\Bzb\ Oscillations with Inclusive Dilepton Events
\end{center}
\bigskip

\begin{center}
\large The \babar\ Collaboration\\
\mbox{ }\\
\today
\end{center}
\bigskip \bigskip

\begin{center}
\large \bf Abstract
\end{center}
We report preliminary results of a search for \CPT and Lorentz violation in 
\Bz -\Bzb 
oscillations using an inclusive dilepton sample collected by the \babar\ 
experiment at the \pep2\ \BF.  Using a sample of 232 million \BB\ pairs, we 
search for time-dependent variations in the complex \CPT parameter 
$\z = \z_0 + \z_1\cos{(\Omega\hat{t} + \phi)}$ where $\Omega$ is the 
Earth's sidereal frequency and $\hat{t}$ is sidereal time.
We measure $\ImzZero = (-14.1 \pm 7.3 (stat.)\pm 2.4(syst.))\times 10^{-3}$,
$\dGRezZero = (-7.2 \pm 4.1(stat.)\pm 2.1(syst.))\times 10^{-3}\ps^{-1}$,
$\ImzOne = (-24.0 \pm 10.7 (stat.)\pm 5.9(syst.))\times 10^{-3}$, and 
$\dGRezOne = (-18.8 \pm 5.5(stat.)\pm 4.0(syst.))\times 10^{-3}\ps^{-1}$, 
where $\Delta\Gamma$ is the difference between the decay rates of the neutral 
$B$ mass eigenstates.
The statistical correlation between the measurements of \ImzZero\ and 
\dGRezZero\ is 76\%; between \ImzOne\ and \dGRezOne\ it is 79\%.
These results are used to evaluate expressions involving coefficients for 
Lorentz and \CPT violation in the general Lorentz-violating standard-model 
extension.  In a complementary approach, we examine the spectral power of 
periodic variations in \z over a wide range of frequencies and find no 
significant signal.

\vfill
\begin{center}

Submitted to the 33$^{\rm rd}$ International Conference on High-Energy Physics, ICHEP 06,\\
26 July---2 August 2006, Moscow, Russia.

\end{center}

\vspace{1.0cm}
\begin{center}
{\em Stanford Linear Accelerator Center, Stanford University, 
Stanford, CA 94309} \\ \vspace{0.1cm}\hrule\vspace{0.1cm}
Work supported in part by Department of Energy contract DE-AC03-76SF00515.
\end{center}

\newpage
}

\input pubboard/authors_ICHEP2006

\section{INTRODUCTION}
\label{sec:Introduction}
It has been shown~\cite{Greenberg_2002} that ``If \CPT invariance is 
violated in an interacting quantum field theory, then that theory also 
violates Lorentz invariance.''
The general Lorentz-violating standard-model extension 
(SME)~\cite{Kostelecky_1} 
has been used to show that the parameter for \CPT violation in neutral meson 
oscillations depends on the 4-velocity of the meson~\cite{Kostelecky_2}.
In studies of \upsbb\ decays at asymmetric-energy $e^+e^-$ colliders, any 
observed \CPT asymmetry should vary with sidereal time as the \FourS boost 
direction rotates together with the Earth~\cite{Kostelecky_3}, completing one 
revolution with respect to the Universe in one sidereal day 
($\approx 0.99727$\,solar day). 
We report a search for such effects using inclusive dilepton events recorded 
by the \babar\ detector at the PEP-II collider.

The physical states of the \Bz-\Bzb\ system are eigenstates of a complex 
$2\times 2$ effective Hamiltonian and may be written as
\begin{eqnarray}
|B_L\rangle&=&p\sqrt{1-\z}|\Bz\rangle +q\sqrt{1+\z}|\Bzb\rangle,  \nonumber \\
|B_H\rangle&=&p\sqrt{1+\z}|\Bz\rangle -q\sqrt{1-\z}|\Bzb\rangle, 
\label{eq:mass_eigenstates_cpt}
\end{eqnarray}
where $L$ and $H$ indicate ``light'' and ``heavy.''
The complex parameter \z vanishes if \CPT is preserved.  \T invariance implies 
$\absqop=1$, and \CP invariance requires $\absqop=1$ and $\z = 0$. 

The leading-order $C\!PT$-violating contributions in the SME imply \z 
depends on the meson 4-velocity $\beta^\mu=\gamma(1,\vec{\beta})$ 
in the observer frame as~\cite{w-xi_formalism}
\begin{eqnarray}
\z \approx \frac{\beta^\mu \Delta a_\mu}{\Delta m - i\Delta\Gamma/2}. 
\label{eq:zBDa}
\end{eqnarray}
Here $\beta^\mu \Delta a_\mu$ is the real part of the difference 
between the diagonal elements of the effective Hamiltonian, and the 
magnitude of the decay rate 
difference $\Delta\Gamma = \Gamma_H - \Gamma_L$ is known to be small compared 
to the \Bz-\Bzb\ oscillation frequency $\Delta m = m_H - m_L$. 
The sidereal time dependence of \z arises from the rotation of $\vec{\beta}$ 
relative to the constant vector $\Delta\vec{a}$.
The $\Delta a_\mu$ contain flavor-dependent $C\!PT$- and Lorentz-violating 
coupling coefficients for the valence quarks in the \Bz\ meson. 
Analogous, but distinct, $\Delta a_\mu$ apply to oscillations of other 
neutral mesons. 
Limits on the $\Delta a_\mu$ specific to \KzKzb\ 
oscillations~\cite{KTEV1} and on the $\Delta a_\mu$ specific to 
\DzDzb\ oscillations~\cite{FOCUS} 
have been reported by the KTeV and FOCUS collaborations, respectively.
KTeV has also reported constraints on sidereal-time variation 
of the \CPT violation parameter $\phi_{+-}$~\cite{KTEV2}.

\begin{figure}[!htb]
\begin{center}
\includegraphics[height=7cm]{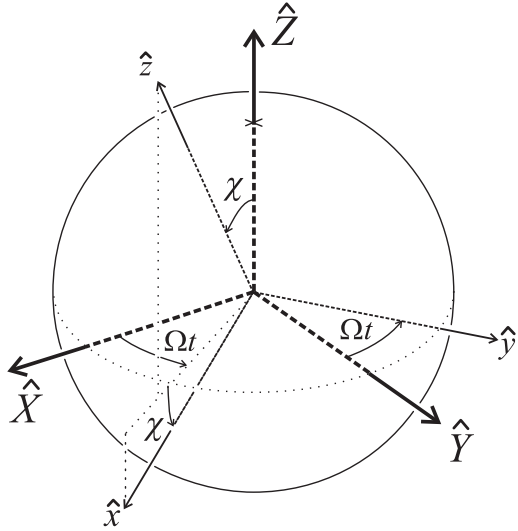}
\caption{Basis $(\hat{x},\hat{y},\hat{z})$ for the rotating laboratory frame,
        and basis $(\hat{X},\hat{Y},\hat{Z})$ for the fixed non-rotating frame.
        The laboratory frame precesses around the Earth's rotation axis 
        $\hat{Z}$ at the sidereal frequency $\Omega$. The angle between 
        $\hat{Z}$ and the direction $\hat{z}$ opposite to the \FourS boost 
        direction at PEP-II is $\chi = 51^\circ$.}
\label{fig:frame_bases}
\end{center}
\end{figure}

We approximate the 4-velocity of each $B$ meson by the \FourS 4-velocity so 
that \z is common to each $B$ in a pair.  We choose the meson 3-velocity to 
lie along  $-\hat{z}$ in the rotating laboratory frame shown in 
Fig.~\ref{fig:frame_bases}. The non-rotating frame containing the constant 
vector $\Delta\vec{a}$ has $\hat{Z}$ along the Earth's rotation axis, 
corresponding to declination $90^\circ$ in celestial equatorial coordinates. 
$\hat{X}$ and $\hat{Y}$, each in the equatorial plane, lie at right ascension 
$0^\circ$ and $90^\circ$, respectively.  At sidereal time $\hat{t}=0$, 
$\hat{z}$ lies in the $\hat{X}$-$\hat{Z}$ plane and $\hat{y}$ is 
coincident with $\hat{Y}$. The \CPT parameter \z may then be expressed as  
\begin{eqnarray}
\z \equiv \z(\hat{t}\,) = {\frac{\gamma}{\Delta m - i\Delta\Gamma/2}}\left[\Delta a_0 - \beta\Delta a_Z\cos\chi - \beta\sin\chi\left(\Delta a_Y\sin\Omega\hat{t} + \Delta a_X\cos\Omega\hat{t}\right)\right], 
\end{eqnarray}
where $\cos\chi = \hat{z}\cdot\hat{Z}$ and 
$\Omega = 2\pi /24$\,rad$\cdot$sidereal-hour$^{-1}$ is the Earth's sidereal 
frequency.

We use the latitude ($37.4^\circ$\,N) and longitude ($122.2^\circ$\,W) of the 
\babar\ detector, together with the Lorentz boost of the \FourS  
($\beta\gamma = 0.55$ directed $37.8^\circ$ east of south), to determine 
$\cos{\chi} = 0.63$ and $\hat{t} = (\hat{t}_G + 7.3$)\,sidereal-hours, where 
$\hat{t}_G$ is Greenwich Mean Sidereal Time (GMST) for each event;  hence 
\begin{equation}
\z(\hat{t}\,) = {\frac{\left[1.14\Delta a_0 - 0.35\Delta a_Z - 0.43\left(\Delta a_Y\sin\Omega\hat{t} + \Delta a_X\cos\Omega\hat{t}\right)\right]}{\Delta m - i\Delta\Gamma/2}}.
\label{eq:zDa0XYZ}
\end{equation}
We convert the ``timestamp'' that records when the event occurred to the 
Julian date (J) and calculate GMST as specified by the U.S.\ Naval 
Observatory~\cite{USNO}:
\begin{eqnarray}
  \hat{t}_G = {\rm mod}(18.697374558 + 24.06570982441908\,{\rm D},\: 24),   
\end{eqnarray}
where ${\rm D} = {\rm J} - 2451545.0$ is the number of Julian days since 
12$^{\rm h}$:\,00 Universal Time on January 1, 2000. 

The clock used to set the event timestamp has a rate governed by the PEP-II 
59.5\,MHz master oscillator and is resynchronized with U.S.\ time 
standards via Network Time Protocol at intervals of less than four months. 
During such intervals the event timestamps are conservatively estimated to 
accumulate absolute errors of less than 30 seconds per month. The sidereal 
phase of each timestamp is therefore determined to better than 0.2\%.

Since sidereal time gains 12\,hours every six months relative to solar time,
possible day/night variations in detector response tend to cancel over 
sidereal time for long data-taking periods. The data used in this analysis 
were accumulated over a period of more than four years.

Inclusive dilepton events, where both $B$ mesons decay semileptonically 
($b \to X\ell\nu$, with $\ell=e$ or $\mu$), comprise 4\% of all \upsbb decays 
and provide a very large data sample for studies of \CPT violation in 
mixing.  In {\it direct} semileptonic neutral $B$ decays, the flavor 
$\Bz(\Bzb)$ is tagged by the charge of the daughter lepton $\ellp(\ellm)$.

At the \FourS resonance, neutral $B$ mesons are produced in a coherent P-wave 
state. The $B$ mesons remain in orthogonal flavor states until one decays, 
after which the flavor of the other $B$ meson continues to evolve in time.
Neglecting second order terms in \z, the decay rates for the three 
semileptonic decay configurations ($\ellp\ellp$, $\ellm\ellm$, $\ellp\ellm$) 
are given by
\begin{eqnarray}
N^{++}\!\! &\propto& \! e^{-\Gamma |\dt|} |p/q|^2\left\{\cosh(\dG \dt /2) - \cos(\dm \dt) \right\}, \nonumber \\
N^{--}\!\! &\propto& \!e^{-\Gamma |\dt|} |q/p|^2\left\{\cosh(\dG \dt /2) - \cos(\dm \dt) \right\}, \nonumber\\
N^{+-}\!\! &\propto& \!e^{-\Gamma |\dt|} \left\{\cosh(\dG \dt /2) 
  - 2\,{\Rez}\sinh(\dG \dt /2) + \cos(\dm \dt) + 2\,{\Imz}\sin(\dm \dt)\right\}, 
\label{eq:decayrate}
\end{eqnarray}
where $\Gamma$ is the average neutral $B$ decay rate, and \dt\ is the 
difference between the proper decay times of the two $B$ mesons.  
The sign of \dt\ has a physical 
meaning only for opposite-sign dileptons and is given by $\dt = t^+ - t^-$, 
where $t^+(t^-)$ corresponds to $\ellp(\ellm)$, respectively.

The opposite-sign dilepton \CPT asymmetry $A_{\CPT}$, between events with 
$\dt >0$ and $\dt<0$, compares the oscillation probabilities $P(\Bz \to \Bz)$ 
and $P(\Bzb \to \Bzb)$ and is sensitive to \CPT violation through the 
parameter \z:
\begin{eqnarray}
A_{\CPT}(|\dt|)& =&  \frac {P(\Bz \to \Bz)-P(\Bzb \to \Bzb)}
                       {P(\Bz \to \Bz) + P(\Bzb \to \Bzb)}
   \;  = \;  \frac {N^{+-}(\dt>0) - N^{+-}(\dt<0)}{N^{+-}(\dt>0) + N^{+-}(\dt<0)}
\nonumber\\
& \simeq & 2\frac{ - {\Rez}\sinh(\dG \dt /2) + {\Imz}\sin(\dm\dt)}
{\cosh(\dG \dt /2) + \cos(\dm\dt)}.
\label{eq:acpt}
\end{eqnarray}
The experimental bound on $|\dG|$~\cite{BaBarCPT} is sufficiently small for 
the approximation $\Rez\sinh(\dG\dt/2) \simeq  \dGRez \times (\dt/2)$ to be 
valid over the range $-15 < \dt < 15$\,ps used in this analysis, and we 
measure the product \dGRez\  instead of \Rez\ alone.

We present measurements of \Imz\ and \dGRez\ using a simultaneous 
two-dimensional likelihood fit to the observed $\dt$ and sidereal time 
$(\hat{t}\,)$ distributions of opposite-sign and same-sign dilepton events. 
Inclusion of the same-sign events allows a better determination of the 
fraction of non-signal events (called ``$obc$'' in Sect.~\ref{sec:Analysis}) 
in which the lepton from one $B$ meson is not a {\it direct} daughter.
We search for variations in \z\ of the form 
\begin{equation}
\z = \z_0 + \z_1\cos{(\Omega{\hat{t}} + \phi)} 
\end{equation}
with a period of one 
sidereal day, and extract values for the $C\!PT$- and Lorentz-violating 
coupling coefficients $\Delta a_\mu$ in the SME from the measured quantities 
\ImzZero, \ImzOne, \dGRezZero, and \dGRezOne.  This extends our previous 
sidereal-time-independent analysis that measured \ImzZero, \dGRezZero, and 
$|q/p|$ with the same events~\cite{thePRL}. 
In the decay rates, we use $|\dG|= 6 \times 10^{-3}\ps^{-1}$ in the 
$\cosh(\dG \dt /2)$ term and $|q/p| = 1$, consistent with the values reported 
in Ref.~\cite{BaBarCPT} and Ref.~\cite{thePRL}, respectively. 
In a complementary approach, we use the periodogram method~\cite{pgram}, 
developed for studies of variable stars, to detect directly any periodic 
variations in \z over a wide range of frequencies and to measure their 
spectral power ${\rm P}(\nu)$.

\section{THE \babar\ DETECTOR AND DATASET}
\label{sec:babar}
This analysis is based on about 232 million \upsbb decays collected during 
1999--2004 with the \babar\ detector at the \pep2\ asymmetric-energy $e^+e^-$ 
storage ring.  An additional 16\,\invfb of ``off-resonance'' data recorded 
40\,\mev below the \FourS is used to model continuum background. 

The \babar\ detector is described in detail elsewhere~\cite{ref:babar}.
This analysis uses the tracking system composed of a five-layer silicon vertex 
tracker (SVT) and a 40-layer drift chamber (DCH), the Cherenkov radiation 
detector (DIRC) for charged $\pi$--$K$ discrimination, the CsI(Tl) calorimeter 
(EMC) for electron identification, and the 18-layer flux return (IFR) located 
outside the 1.5-T solenoid coil and instrumented with resistive-plate 
chambers for muon identification and hadron rejection.
A detailed Monte Carlo program based on {\mbox{\tt GEANT4}}~\cite{geant4} is 
used to simulate the response and performance of the \babar\ detector.

\section{ANALYSIS METHOD AND LIKELIHOOD FIT}
\label{sec:Analysis}
The event selection is similar to that described in
Ref.~\cite{BaBarAT}. Non-\BB background, mainly due to $e^+e^- \to
q\overline{q}$ $(q = u, d, s, c)$ continuum  events, is suppressed
by applying requirements on the shape and the topology of the
event.

Lepton candidate tracks must  have at least 12 hits in the DCH, at least one 
$z$-coordinate measurement in the SVT, and momentum between 0.8 and 2.3\,\gevc 
in the \FourS rest frame.
Electrons are selected by requirements on the ratio of the energy deposited in 
the EMC to the momentum measured in the DCH.
Muons are identified through the energy released in the EMC, as well as the 
strip multiplicity, track continuity, and penetration depth in the IFR. 
Lepton candidates are rejected if their signal in the DIRC is consistent with 
that of a kaon or a proton. 
The electron and muon selection efficiencies are about 85\% and 55\%, with 
pion misidentification probabilities around 0.2\% and 3\%, respectively.

Electrons from photon conversions are identified and rejected with a 
negligible loss of efficiency for signal events.
Leptons from \jpsi\ and $\psi (2S)$ decays are identified by pairing them with 
other oppositely-charged candidates of the same lepton species, selected with 
looser criteria. The event is rejected if the invariant mass of any such 
lepton pair satisfies $3.037 < m_{\ell^+\ell^-} <3.137$\,GeV/$c^2$ or 
$3.646 < m_{\ell^+\ell^-} <3.726$\,GeV/$c^2$.
Remaining events with at least two leptons are retained, and the two highest 
momentum leptons in the \FourS rest frame are used as the dilepton candidates.

Separation between {\it direct} leptons (``$\;b \to \ell\;\,$'') and 
background {\it cascade} leptons from the 
``$\;b\rightarrow c\rightarrow \ell\;\,$'' decay 
chain is achieved with a neural network that combines five discriminating 
variables: the momenta of the two lepton candidates, the angle between the 
momentum directions of the two leptons, and the total visible energy and 
missing momentum in the event, all computed in the \FourS\ rest frame.
Of the original sample of 232 million \BB\ pairs, 1.4 million pass the 
selection.

In the inclusive approach used here, the $z$ coordinate of the $B$ decay point
is approximated by the $z$ coordinate of the lepton candidate's 
point of closest approach in the transverse plane to our best estimate of the 
$(x,y)$ decay point of the \FourS.
In the transverse plane, both the intersection point of the lepton tracks and 
the beam-spot position provide information about the \FourS decay point. 
We combine this information in a $\chi^2$-fit that optimizes our estimate of 
the \FourS decay point in the transverse plane using the transverse distances 
to the two lepton tracks and the transverse distance to the beam-spot 
position. 
The proper time difference $\dt$ between the two $B$ meson decays is taken as 
$\dt=\Delta z/ \langle\beta\gamma\rangle c$, where $\Delta z$ is the 
difference between the $z$ coordinates of the $B$ decay points, with the same 
sign convention as for \dt, and $\langle \beta \gamma \rangle = 0.55$ is the 
nominal Lorentz boost.
For same-sign dileptons, the sign of \dt\ is chosen randomly.

A large control sample of $e^+e^-\rightarrow \mu^+\mu^-(\gamma)$ events, with 
true $\Delta z = 0$, was used to check for any sidereal-time-dependent bias 
in the $\Delta z$ measurement that could mimic a signal for Lorentz violation.
The measured amplitude for such a bias at the sidereal 
frequency is $(0.015\pm 0.025)\,\mu$m, consistent with no variation
around the mean value $\langle \Delta z \rangle = (0.030\pm 0.018)\,\mu$m.
The corresponding amplitude for a sidereal-time-dependent bias in \dt\ for 
\BB\ events is $(9\pm 15)\times 10^{-5}$\,ps. Similar amplitudes are found 
for possible day/night variations in the $\Delta z$ and \dt\ measurements.

We model the contributions to our sample from \BB\ decays using five 
categories of events, $i$, each represented by a probability density function 
(PDF) in $\dt$ and sidereal time $\hat{t}$, denoted by ${\cal P}_{i}^{n,c}$. 
Their shapes  are determined using the $\BzBzb$ $(n)$ and $\BpBm$ $(c)$ Monte 
Carlo simulation separately, with the approach described in 
Ref.~\cite{BaBardm}.

The five categories of dilepton \BB\ decays, with contributions 
estimated from Monte Carlo simulation, are the following.
Pure signal events with two direct leptons ($sig$), comprising 81\% 
of the selected \BB\ events, give information about the \CPT parameter \z. 
``Opposite $B$ cascade'' ($obc$) events, where the direct lepton and the 
cascade lepton come from different $B$ decays, contribute about 9\%.
``Same $B$ cascade'' ($sbc$) events, in which the direct lepton and 
the cascade lepton stem from the same $B$ decay, contribute around 4\%.  
About 3\% of the dilepton events originate from the decay chain 
``$\;b \to \tau^- \to \ell^-\;$'' ($1d1\tau$), which tags the $B$ flavor 
correctly.  The remaining \BB\ events ($other$) consist mainly of one direct 
lepton and one lepton from the decay of a charmonium resonance from the other 
$B$ decay.

The signal event PDFs, ${\cal P}_{sig}^{n,c}$, are the convolution of an 
oscillatory term containing the sidereal-time dependent \CPT parameter 
(Eq.~\ref{eq:decayrate}) for neutral $B$ decays (or an exponential function 
for charged $B$ decays) with a resolution function that is the sum of three 
Gaussians (core, tail, and outlier) with means fixed to 
zero~\cite{bib:BABAR-s2b}. The widths of the narrower core and tail Gaussians 
are free parameters in the fit to data; the width of the outlier Gaussian is 
fixed to $8\ps$. The fractions of all three Gaussians are determined by the 
fit: the tail and outlier fractions are free parameters, and the sum of the 
three fractions is constrained to unity.

The $obc$ event PDFs,  ${\cal P}_{obc}^{n,c}$, are modeled by the convolution 
of ($\dt, \hat{t}\,)$-dependent terms, similar in form to those for signal, 
with a resolution function that takes into account the effect of the charmed 
meson lifetimes.
Since both short-lived $D^0$ and \Ds, and long-lived $D^+$ mesons are involved 
in cascade decays, the resolution function for the long-lived and short-lived 
components is the convolution of a double-sided exponential with the sum of 
three Gaussians.
To allow for possible outliers not present in the Monte Carlo simulation, the
fraction of the outlier Gaussian is a free parameter in the fit to data. 
The parameterization of the $sbc$ event PDFs, ${\cal P}_{sbc}^{n,c}$, account 
for the lifetimes of charmed mesons in a similar way.

The PDFs for $1d1\tau$ events, ${\cal P}_{1d1\tau}^{n,c}$, are similar to 
those for the signal events.
The resolution function takes into account the $\tau$ lifetime and is 
chosen to be the convolution of two double-sided exponentials with two 
Gaussians.
The PDFs for the remaining \BB\ events, ${\cal P}_{other}^{n,c}$, are the 
convolution  of an exponential function with an effective lifetime and two 
Gaussians.

The fractions ($f_{sbc}^{n,c}$, $f_{1d1\tau}^{n,c}$ and $f_{other}^{n,c}$) of 
$sbc$, $1d1\tau$ and $other$ events are determined directly from the $\BzBzb$ 
and $\BpBm$ Monte Carlo simulations. 
The fraction $f_{+-}$ of  $\BpBm$ events and the fraction $f_{obc}^{n}$ of 
\BzBzb $obc$ events are free parameters in the fit to data. 
The ratio $f_{obc}^{c}/f_{obc}^{n}$ is constrained to the estimate obtained 
from Monte Carlo samples. 

Non-\BB\ events are estimated, using off-resonance data, to comprise 
$f_{cont}= (3.1\pm0.1)\%$ of the dilepton candidates.  The PDF for this 
component is modeled using off-resonance dilepton candidates selected with 
looser criteria and on-resonance events that fail the continuum-rejection 
criteria.

The \CPT violation parameter \z 
is extracted from a binned maximum likelihood fit to the events that pass the 
dilepton selection. The likelihood ${\cal L}$ contains the 
$(\dt,\hat{t}\,)$-dependent PDFs described previously and combines 24 
sidereal-time bins. 

The likelihood for each sidereal-time bin is given by
\begin{eqnarray}
  { \cal L}(\dt)   & = & f_{cont} {\cal P}_{cont} + (1-f_{cont}) \{  f_{+-}{\cal P}_{B^+B^-}
  + (1-f_{+-}){\cal P}_{B^0\overline B^0} \}, \,\,\,{\rm where} \nonumber \\
  {\cal P}_{\BzBzb} & = & (1-f^n_{sig}){\cal P}^n_{casc} + f^n_{sig}{\cal P}^n_{sig}, \nonumber\\
  {\cal P}_{\BpBm} & = & (1-f^c_{sig}){\cal P}^c_{casc} + f^c_{sig}{\cal P}^c_{sig}, \nonumber \\
  {\cal P}^{n,c}_{casc} & = & f^{n,c}_{other} {\cal P}^{n,c}_{other} + f^{n,c}_{1d1\tau} {\cal P}^{n,c}_{1d1\tau}
  + f^{n,c}_{sbc} {\cal P}^{n,c}_{sbc} + f^{n,c}_{obc} {\cal P}^{n,c}_{obc}. 
\end{eqnarray}
Here we omit small charge asymmetries ($\sim 10^{-3}$) induced by 
charge-dependent lepton reconstruction and identification efficiencies. 
While affecting the same-sign decay rates, these asymmetries cancel at first 
order for opposite-sign dilepton events.
Our previous sidereal-time-independent analysis~\cite{thePRL} found these 
asymmetries to be a source of systematic uncertainty only for $|q/p|$. 

The likelihood fit gives 
$\ImzZero = (-14.1 \pm 7.3)\times 10^{-3}$,
$\dGRezZero = (-7.2 \pm 4.1)\times 10^{-3}\ps^{-1}$,
$\ImzOne = (-24.0 \pm 10.7)\times 10^{-3}$, and 
$\dGRezOne = (-18.8 \pm 5.5)\times 10^{-3}\ps^{-1}$.
The statistical correlation between the measurements of \ImzZero\ and 
\dGRezZero\ is 76\%; between the measurements of \ImzOne\ and \dGRezOne\ it is 
79\%.
The fitted fractions of $\BpBm$ and $obc$ events are
$f_{+-}=(59.1\pm0.3)\%$ and $f^{n}_{obc}= (10.7\pm0.1)\%$, respectively.

\begin{figure}[!htb]
\begin{center}
\includegraphics[height=7.3cm]{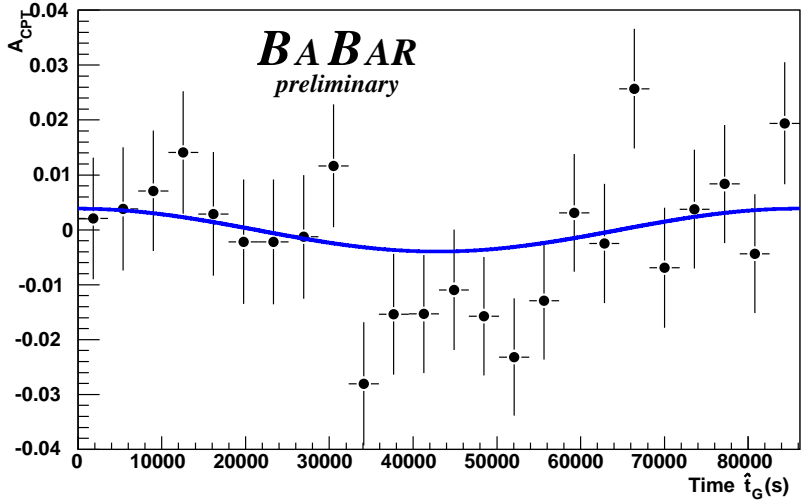}
\caption{The asymmetry $A_{\CPT}$, integrated over $|\dt| > 3\ps$, as a 
         function of Greenwich Mean Sidereal Time in seconds folded over a 
         period of 24 sidereal hours. The curve is a 
         projection from the two-dimensional likelihood 
         fit onto the sidereal time axis.}
\label{fig:CPTAsyData1}
\end{center}
\end{figure}

Figure~\ref{fig:CPTAsyData1} shows the asymmetry $A_{\CPT}$, defined in 
Eq.~\ref{eq:acpt}, as a function of sidereal time. The curve is a projection 
of the two-dimensional asymmetry onto the sidereal-time axis.
To exhibit better the measured asymmetry, the projection is performed by   
integrating over $|\dt| > 3\ps$ thereby omitting highly-populated bins near 
$|\dt| = 0$ where any asymmetry is predicted to be small and is diluted by 
\dt\ resolution effects.

\section{SYSTEMATIC STUDIES}
\label{sec:Systematics}

There are several sources of systematic uncertainty in these measurements.
To determine their magnitude, we vary each source of systematic effect by its 
known or estimated uncertainty, and take the resulting deviation in each of 
the measured parameters as its error.

The widths of the core and tail Gaussians of the resolution function for 
the $obc$ and $sbc$ categories as well as the pseudo-lifetime for the 
$1d1\tau$ and $other$ categories are varied separately by 10\%. 
This variation is motivated by comparing the fitted parameters of the 
signal resolution function obtained from \BB\ Monte Carlo samples
and from data. The fractions of the short-lived and long-lived
charmed meson components  for $obc$ and $sbc$ are varied by 10\%. 
Modeling of the PDFs is the main source of systematic uncertainty in \ImzZero.

We have also varied the parameters  $\dm$, $\tau_{\Bz}$ and
$\tau_{\Bpm}$ independently within their known
uncertainties~\cite{bib:HFAG}. 
For \dG, we have allowed for $3\sigma$ deviations from the value reported in 
Ref.~\cite{BaBarCPT} by varying $|\dG|$ over the range 0\,--\,0.1\,ps$^{-1}$.
The lifetimes $\tau_{\Bz}$ and $\tau_{\Bpm}$ are the dominant sources of 
systematic uncertainty in \ImzOne\ and \dGRezOne. 
The dominant systematic uncertainty in \dGRezZero\ is imperfect knowledge of 
the absolute $z$ scale of the detector and residual uncertainties in the SVT 
local alignment. A possible sidereal-time-dependent bias in the \dt\ 
measurement with amplitude $24\times 10^{-5}$\,ps, derived from the amplitude 
$(9\pm 15)\times 10^{-5}$\,ps found with 
$e^+e^-\rightarrow \mu^+\mu^-(\gamma)$ events, contributes a negligible 
systematic uncertainty.

\begin{table} [!htb]
\caption{Summary of systematic uncertainties  for \ImzZero, \dGRezZero,
             \ImzOne, and \dGRezOne.}
\begin{center}
\begin{tabular}{lcccc}
\hline
\hline
{\bf Systematic Effects}                    & $\sigma(\ImzZero)$ & $\sigma(\dGRezZero)$       & $\sigma(\ImzOne)$  & $\sigma(\dGRezOne)$\\
                                            & $(\times 10^{-3})$ & $(\times 10^{-3}\ps^{-1})$ & $(\times 10^{-3})$ & $(\times 10^{-3}\ps^{-1})$\\
\hline
PDF modeling                                &  $\pm 2.0$         &  $\pm 1.0$                 &  $\pm 2.5$         &  $\pm 1.2$  \\
Bkgd component\ fractions                   &  $\pm 0.1$         &  $\pm 0.1$                 &  $\pm 0.2$         &  $\pm 0.2$  \\
\dG, $\dm$, $\tau_{\Bz}$, $\tau_{\Bpm}$     &  $\pm 1.3$         &  $\pm 1.0$                 &  $\pm 4.9$         &  $\pm 3.6$  \\
SVT alignment                               &  $\pm 0.6$         &  $\pm 1.5$                 &  $\pm 2.0$         &  $\pm 1.1$  \\
\hline
Total                                       &  $\pm 2.4$         &  $\pm 2.1$                 &  $\pm 5.9$         &  $\pm 4.0$  \\
\hline
\hline
\end{tabular}
\label{tab:Syst}
\end{center}
\end{table}

For each parameter, the total
systematic uncertainty is the sum in quadrature of the estimated systematic
uncertainties from each source, as summarized in Table~\ref{tab:Syst}.

\section{\boldmath \CPT VIOLATION PARAMETER FREQUENCY ANALYSIS}
\label{sec:Frequency}
To perform a more general search for periodic variations in the \CPT violation 
parameter \z\ over a wide frequency range, we adopt the periodogram 
method~\cite{pgram} used in astronomy to study the periodicity of variable 
stars such as Cepheids. For each test frequency $\nu$, the method 
determines the spectral power ${\rm P}(\nu)$, defined by
\begin{equation}
{\rm P}(\nu) \equiv \frac{1}{N\sigma^2_w}\Bigl|\sum_{j=1}^{N}w_j e^{2i\pi\nu T_j}\Bigr|^2,
\end{equation}
from $N$ data points measured at times $T_j$ and having weights $w_j$ with 
variance $\sigma_w^2$. In our case $T_j$ is the Universal Time of event $j$. 
In the absence of an oscillatory signal, the probability 
that the largest ${\rm P}(\nu)$ exceeds a value $S$ is given by
\begin{equation}
\Pr\big\{ {\rm P_{max}}(\nu)>S;M\big\} = 1 - \Bigl(1 - e^{-S}\Bigr)^M,
\label{eq:peri-pr}
\end{equation}
where $M$ is the number of independent frequencies tested.

\begin{figure}[!htb]
\begin{center}
\includegraphics[width=8.3cm]{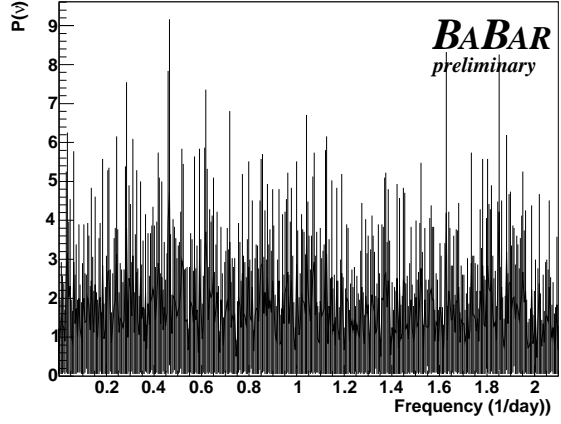} \\
\includegraphics[width=8.3cm]{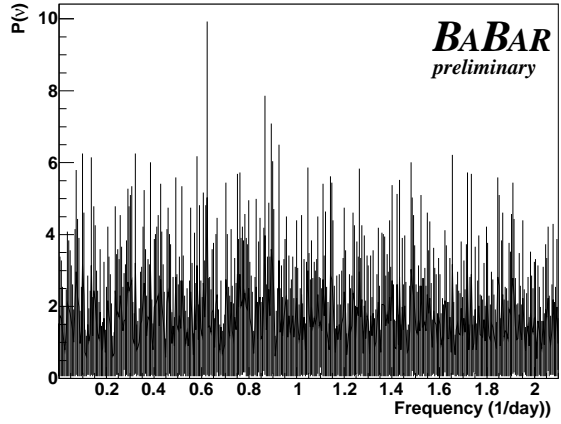}    \\
\includegraphics[width=8.3cm]{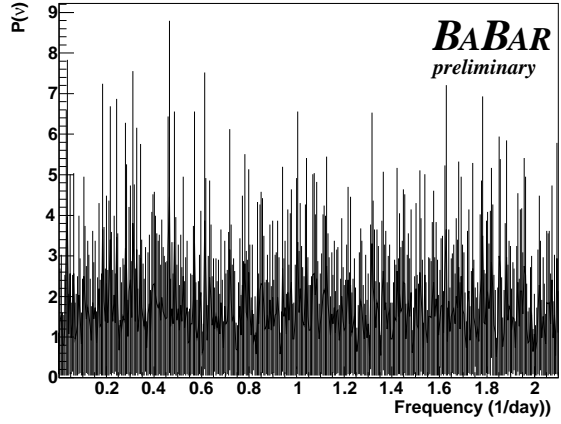}
\caption{Periodograms for opposite-sign dileptons showing spectral power 
         P$(\nu)$ for weights 
         $w_j \propto \dt_j$ (top), $w_j \propto \sin{(\dm\dt_j)}$ (center), 
         and $w_j\propto \dm\dt_j - \sin(\dm\dt_j)$ (bottom) providing 
         sensitivity to \dGRezOne, to \ImzOne, and to \ImzOne\ subject to the
         SME constraint ${\dGRez} = 2\dm\,\Imz$, respectively.}
\label{fig:periodogram}
\end{center}
\end{figure}

Our search uses 20994 test frequencies from 0.26\,year$^{-1}$ to 
2.1\,day$^{-1}$ in units of (solar time)$^{-1}$, with steps of 
$10^{-4}$\,day$^{-1}$. To guard against 
underestimating the spectral power of a signal, we have oversampled the 
frequency range by a factor of about 2.2. The number of independent 
frequencies is about 9500. 
Twenty-seven test frequencies lie between the Earth's sidereal 
and solar rotation frequencies. Each weight $w_j$ depends on the decay time 
difference $\dt_j$ reconstructed for event $j$ occuring at time $T_j$.
Periodic variations in \z\ affect the decay rate $N^{+-}$ through the terms 
${\Imz}\sin(\dm \dt)$ and ${\Rez}\sinh(\dG\dt/2)\simeq  \dGRez\,(\dt/2)$ in 
Eq.~\ref{eq:decayrate}.
Sensitivity to variations in \dGRez\ and \Imz\ is attained by employing 
weights $w_j\propto \dt_j$ and $w_j\propto \sin(\dm\dt_j)$, respectively.  
In the context of the SME, the imaginary part of 
Eq.~\ref{eq:zBDa} implies 
${\dGRez} = 2\dm\,\Imz$, and hence in Eq.~\ref{eq:decayrate} we have
${\Imz}\,\sin{(\dm\dt)} - {\Rez}\,\sinh{(\dG\dt /2)}\; \simeq \; {\Imz}\left[\,\sin{(\dm\dt)} - \dm\dt\,\right]$.
Accordingly, we also search for periodic variations in \Imz\ using 
weights $w_j\propto \dm\dt_j - \sin(\dm\dt_j)$.

Figure~\ref{fig:periodogram} shows the spectral powers ${\rm P}(\nu)$ 
measured in the opposite-sign dilepton data sample using the weights $\dt_j$, 
$\sin{(\dm\dt_j)}$, and $\dm\dt_j - \sin(\dm\dt_j)$.
The largest spectral power obtained for each of these weights corresponds to 
statistical fluctuation probabilities of 62\%, 36\%, and 76\%, 
respectively, consistent with no periodic variation in the \CPT violation 
parameter over the frequency range 0.26\,year$^{-1}$ to 2.1\,day$^{-1}$.
At the Earth's sidereal frequency ($\approx 1.0027\,{\rm day}^{-1}$),
${\rm P}(\nu) = 3.73$, 0.71, and 6.24 for the three weight types. 
At the Earth's solar-day frequency, the corresponding ${\rm P}(\nu) = 1.50$,
0.97, and 1.47.

To check the validity of these results, we performed several tests 
of the periodogram method using events from data and from Monte Carlo 
simulation. Test periodograms showed large spectral powers at  
expected frequencies for (i) a generic dilepton Monte Carlo sample assigned 
event times $T_j$ with a 0.5\,sidereal-day$^{-1}$ frequency modulation, and 
(ii) unweighted dilepton data events, which give sensitivity to variations 
in the overall event rate --- generally higher during night and weekend 
shifts, corresponding to frequencies of 1\,day$^{-1}$ and 1\,week$^{-1}$.
Test periodograms for same-sign dilepton data events, which are not sensitive 
to \CPT violation, showed no significant spectral power. 
The largest ${\rm P}(\nu)$ value, obtained with $\dt_j$ weights, corresponds 
to a statistical fluctuation probability of 7\%. 
Test periodograms for opposite-sign dilepton data events, with the 
sign of \dt\ randomized to remove any measurable \CPT violation, also showed 
no significant spectral power. We used these periodograms to check whether the 
distribution of ${\rm P}(\nu)$ values has a probability 
density $\propto \exp\{-k\cdot{\rm P}(\nu)\}$ with $k=1$, consistent with 
Eq.~\ref{eq:peri-pr}. A fit to the ${\rm P}(\nu)$ values yields 
$k = 1.006\pm 0.001$ with $\chi^2 = 68.2$ for 53 degrees of freedom.

\section{RESULTS}
\label{sec:Results}
\begin{figure}[!htb]
\begin{center}
\includegraphics[height=9cm]{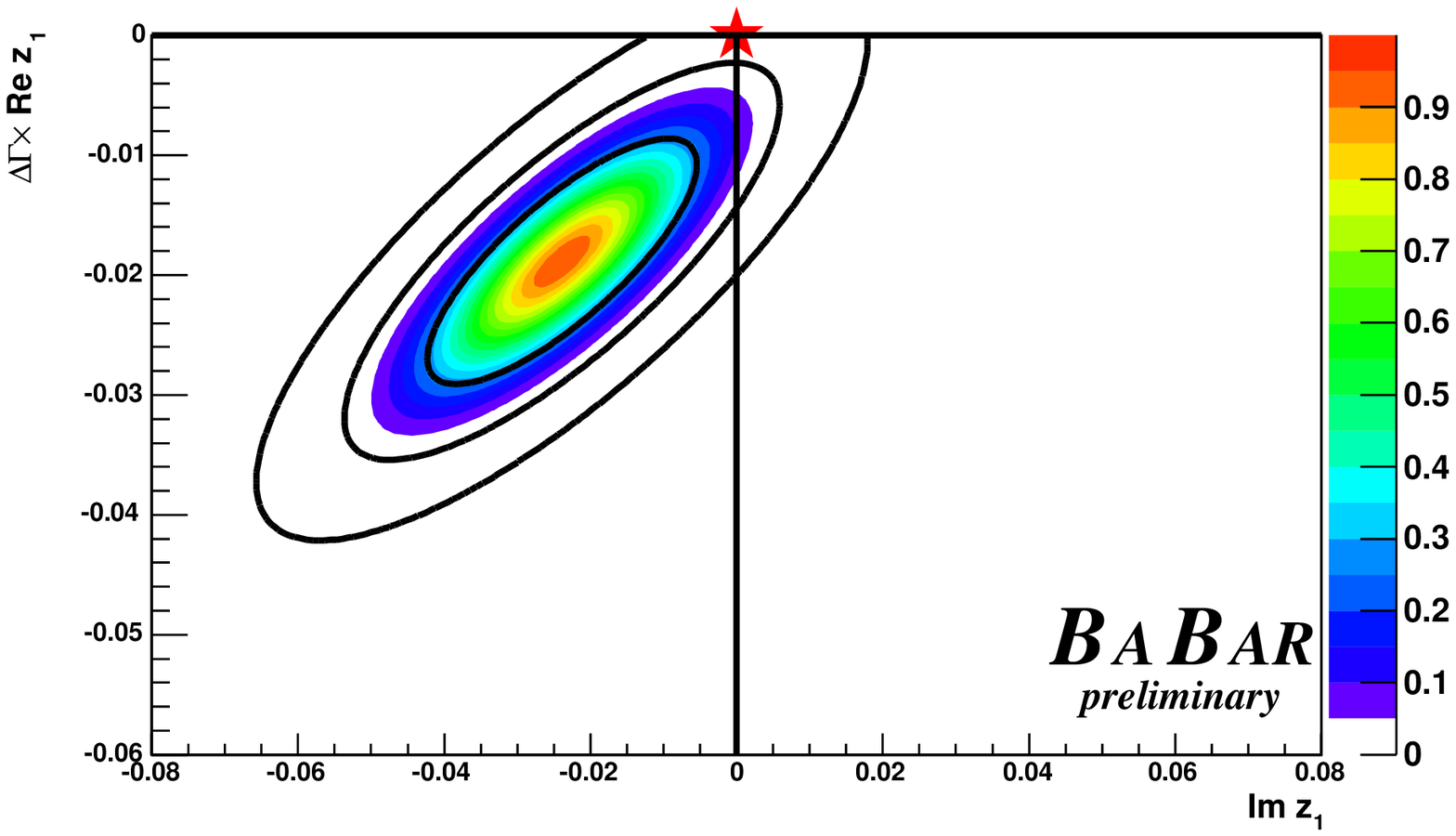}
\caption{Confidence level contours for the parameters \ImzOne\ and \dGRezOne\ 
         including both statistical and systematic errors. The correlation 
         between the measurements of \ImzOne\ and \dGRezOne\ is 79\%.
         The line contours indicate $1\sigma$, $2\sigma$, and $3\sigma$ 
         significance. The star at $\ImzOne = \dGRezOne = 0$ indicates the 
         condition for no sidereal-time dependence in \z.}
\label{fig:CPTAsyData2}
\end{center}
\end{figure}

Figure~\ref{fig:CPTAsyData2} shows confidence level contours for the  
parameters \ImzOne\ and \dGRezOne\ including both statistical and systematic 
errors.
A significance of $2.2\sigma$ is found for periodic variations in the \CPT 
violation parameter \z\ at the sidereal frequency, characteristic of 
Lorentz violation. 

In the framework of the SME, the quantities \ImzZero, \ImzOne, 
\dGRezZero, and \dGRezOne\ are related by Eq.~\ref{eq:zDa0XYZ} to the 
$\Delta a_\mu$ containing $C\!PT$- and Lorentz-violating coupling 
coefficients. With $|\Delta\Gamma| \ll \Delta m$, and using the SME constraint 
${\dGRez} = 2\dm\,\Imz$ implied by Eq.~\ref{eq:zBDa}, we obtain
\begin{equation}
\renewcommand{\arraystretch}{2}
\begin{array}{rcccl}
1.14\Delta a_0 - 0.35\Delta a_Z & \approx & (\Delta m/\Delta\Gamma)\dGRezZero 
       & = & 2\Delta m (\Delta m/\Delta\Gamma)\ImzZero,  \\
0.43^2\left[(\Delta a_X)^2 + (\Delta a_Y)^2\right] & \approx & \left[(\Delta m/\Delta\Gamma)\dGRezOne\right]^2 & = & 4\left[\Delta m (\Delta m/\Delta\Gamma)\ImzOne\right]^2.
\end{array}
\renewcommand{\arraystretch}{1}
\end{equation}
Taking into account error correlations between \dGRezZero\ and \ImzZero, and 
between \dGRezOne\ and \ImzOne, we find
\begin{eqnarray*}
\Delta a_0 - 0.30\Delta a_Z & \approx & -(5.2\pm 4.0)(\Delta m/\Delta\Gamma)\times 10^{-15}\,{\rm GeV},  \\ [0.1in]
\sqrt{(\Delta a_X)^2 + (\Delta a_Y)^2} & \approx & (37\pm 16)|\Delta m/\Delta\Gamma|\times 10^{-15}\,{\rm GeV}. 
\end{eqnarray*} 
Here we use $\Delta m = (0.507\pm 0.004)\ps^{-1} = 
(3.34\pm 0.03)\times 10^{-13}$\,GeV~\cite{PDG2006}, and note that lattice QCD 
calculations give $\Delta m/\Delta\Gamma \sim -200$ in the standard 
model~\cite{Schneider}.

\section{CONCLUSIONS}
\label{sec:Summary}
We have used data containing 232 million \BB\ pairs 
to perform a simultaneous likelihood fit of same-sign and
opposite-sign dilepton events that includes both the decay time 
difference \dt\ and the sidereal time $\hat{t}$ of each 
event. We have measured the \CPT violation parameter of form 
$\z = \z_0 + \z_1\cos{(\Omega{\hat{t}} + \phi)}$ and find
\begin{eqnarray*}
\ImzZero   & = & (-14.1  \pm 7.3  {\rm (stat.)} \pm 2.4{\rm (syst.)})\times 10^{-3},\\
\dGRezZero & = & (-7.2   \pm 4.1  {\rm (stat.)} \pm 2.1{\rm (syst.)})\times 10^{-3}\ps^{-1},\\
\ImzOne    & = & (-24.0   \pm 10.7 {\rm (stat.)} \pm 5.9{\rm (syst.)})\times 10^{-3},\\
\dGRezOne  & = & (-18.8   \pm 5.5  {\rm (stat.)} \pm 4.0{\rm (syst.)})\times 10^{-3}\ps^{-1}.
\end{eqnarray*}
A significance of $2.2\sigma$, compatible with no sidereal-time dependence, is 
found for periodic variations in \z\ at the sidereal frequency that are 
characteristic of Lorentz violation.
The complementary periodogram method provides no strong evidence 
for Lorentz and \CPT violation, or for any periodicity in \z\ over 
the frequency range 0.26\,year$^{-1}$ to 2.1\,day$^{-1}$.
The results of the likelihood fit are used to constrain the quantities 
$\Delta a_\mu$ containing $C\!PT$- and Lorentz-violating coupling coefficients 
for neutral $B$ oscillations in the general Lorentz-violating standard-model 
extension.

\section{ACKNOWLEDGMENTS}
\label{sec:Acknowledgments}
The authors wish to thank V.\ Alan Kosteleck$\acute{\rm y}$ for his advice on
interpreting the measured quantities in the standard-model extension and for 
providing Fig.~\ref{fig:frame_bases}, and Alain Milsztajn for his help with 
the periodogram analysis.
\input pubboard/acknowledgements

\end{document}

%% file: pubboard/authors_ICHEP2006.tex
\begin{center}
\small

The \babar\ Collaboration,
\bigskip

%
{B.~Aubert,}
{R.~Barate,}
{M.~Bona,}
{D.~Boutigny,}
{F.~Couderc,}
{Y.~Karyotakis,}
{J.~P.~Lees,}
{V.~Poireau,}
{V.~Tisserand,}
{A.~Zghiche}
\inst{Laboratoire de Physique des Particules, IN2P3/CNRS et Universit\'e de Savoie,
 F-74941 Annecy-Le-Vieux, France }
% ...

\end{center}\newpage

%% file: pubboard/acknowledgements.tex
We are grateful for the 
extraordinary contributions of our \pep2\ colleagues in
achieving the excellent luminosity and machine conditions
that have made this work possible.
The success of this project also relies critically on the 
expertise and dedication of the computing organizations that 
support \babar.
The collaborating institutions wish to thank 
SLAC for its support and the kind hospitality extended to them. 
This work is supported by the
US Department of Energy
and National Science Foundation, the
Natural Sciences and Engineering Research Council (Canada),
Institute of High Energy Physics (China), the
Commissariat \`a l'Energie Atomique and
Institut National de Physique Nucl\'eaire et de Physique des Particules
(France), the
Bundesministerium f\"ur Bildung und Forschung and
Deutsche Forschungsgemeinschaft
(Germany), the
Istituto Nazionale di Fisica Nucleare (Italy),
the Foundation for Fundamental Research on Matter (The Netherlands),
the Research Council of Norway, the
Ministry of Science and Technology of the Russian Federation, 
Ministerio de Educaci\'on y Ciencia (Spain), and the
Particle Physics and Astronomy Research Council (United Kingdom). 
Individuals have received support from 
the Marie-Curie IEF program (European Union) and
the A. P. Sloan Foundation.

%% file: conf_06_020_copy.bbl
\begin{thebibliography}{99}

\bibitem{Greenberg_2002}
O.~W.~Greenberg, Phys.\ Rev.\ Lett.\ {\bf 89}, 231602 (2002).

\bibitem{Kostelecky_1}
D.~Colladay and V.~A.~Kosteleck$\acute{\rm y}$, Phys.\ Rev.\ D {\bf 55}, 6760
(1997); Phys.\ Rev.\ D {\bf 58}, 116002 (1998); 
V.~A.~Kosteleck$\acute{\rm y}$, Phys.\ Rev.\ D {\bf 69}, 105009 (2004).

\bibitem{Kostelecky_2}
V.~Alan Kosteleck$\acute{\rm y}$, Phys.\ Rev.\ Lett.\ {\bf 80}, 1818 (1998).

\bibitem{Kostelecky_3}
V.~Alan Kosteleck$\acute{\rm y}$, Phys.\ Rev.\ D {\bf 64}, 076001 (2001).

\bibitem{w-xi_formalism}
Ref.~\cite{Kostelecky_2} uses the ``{\sl w}$\xi$'' formalism. In our notation 
$\z = -\xi$ and $|q/p| = $\,{\sl w}.

\bibitem{KTEV1}
KTeV Collaboration, H.\ Nguyen,
in V.~A.\ Kosteleck$\acute{\rm y}$, ed., {\it CPT and Lorentz Symmetry II},
World Scientific, Singapore, 2002.

\bibitem{FOCUS}
FOCUS Collaboration, J.~M.~Link {\it et al.}, Phys.\ Lett.\ B {\bf 556}, 7 (2003).

\bibitem{KTEV2}
KTeV Collaboration, Y.~B.~Hsiung, Nucl.\ Phys.\ B (Proc.\ Suppl.) {\bf 86}, 312 (2000).

\bibitem{USNO} http://aa.usno.navy.mil/faq/docs/GAST.html gives equations for 
GMST adapted from Appendix A of USNO Circular No.\ 163 (1981).

\bibitem{BaBarCPT} \babar\ Collaboration,  B.\ Aubert {\it et al.},
                   \jprd{\bf 70}, 012007 (2004).

\bibitem{thePRL} \babar\ Collaboration, B.\ Aubert {\it et al.}, 
                 \jprl{\bf 96}, 251802 (2006).

\bibitem{pgram} N.~R.~Lomb, Astrophys.\ Space Sci., {\bf 39}, 447 (1976); 
                J.~D.~Scargle, Astrophys.\ J, {\bf 263}, 835 (1982).

\bibitem{ref:babar} \babar\ Collaboration, B.\ Aubert {\it et al.}, Nucl.\ Instrum.\ Methods {\bf A479}, 1 (2002).

\bibitem{geant4} S.~Agostinelli {\it et al.}, Nucl.\ Instrum.\ Methods {\bf A506}, 250 (2003).

\bibitem{BaBarAT} \babar\ Collaboration, B.\ Aubert {\it et al.},
                  \jprl{\bf 88}, 231801 (2002).

\bibitem{BaBardm} \babar\ Collaboration, B.\ Aubert {\it et al.},
                  \jprl{\bf 88}, 221803 (2002).

\bibitem{bib:BABAR-s2b} \babar\ Collaboration, B.\ Aubert {\it et al.},
                        \jprd{\bf 66}, 032003 (2002).

\bibitem{bib:HFAG} Heavy Flavor Averaging Group, K.\ Anikeev {\it et al.}, hep-ex/0505100.

\bibitem{PDG2006}
2006 Review of Particle Physics, W.-M.~Yao {\it et al.}, Journal of Physics G {\bf 33}, 1 (2006).

\bibitem{Schneider} O.~Schneider,``$B^0$-$\overline{B}^0$ Mixing'' in W.-M.~Yao {\it et al.}, Journal of Physics G {\bf 33}, 1 (2006).

\end{thebibliography}
